\documentclass[aps, prb, a4paper, notitlepage, reprint, amsfonts, amsmath, amssymb,floatfix]{revtex4-1}

\usepackage{amsmath}
\usepackage[utf8]{inputenc}
\usepackage{graphicx}				
\usepackage{epstopdf}
\usepackage{xcolor}
\usepackage{physics}
\usepackage{float}
\usepackage{hyperref}

\begin{document}

\title{Luttinger liquid properties from tensor network data}
\author{Michael Weyrauch$^1$}
\author{Mykhailo V. Rakov$^{1,2,3}$}
\affiliation{Physikalisch-Technische Bundesanstalt, Bundesallee 100, D-38116 Braunschweig, Germany}
\affiliation{Technische Universit\"at Braunschweig, Mendelssohnstra\ss e 3, D-38106 Braunschweig, Germany}
\affiliation{Department of Physics and Astronomy, Uppsala University, Box 516, S-75120 Uppsala, Sweden}
\date{\today}

\begin{abstract}
We study the XXZ Heisenberg model in a staggered magnetic field using the HOTRG tensor renormalization
method. Built into the tensor representation of the XXZ model is the U(1) symmetry,
which is systematically maintained at each renormalization step.
We determine the phase diagram of the model from the low lying spectrum, and
from the finite size dependence of the spectrum we extract scaling dimensions, which are compared to
predictions of low energy field theory.
\end{abstract}
\maketitle

\section{Introduction}

The (1+1)-dimensional spin-1/2 XXZ Heisenberg model in a staggered magnetic field features an extended
massless spin-liquid phase. In addition, there are two massive phases, a ferromagnetic phase (FM) and an anti-ferromagnetic phase (AFM). The massless phase is not polarized but ordered partially
anti-ferromagnetically.
A sketch of the phase diagram based on calculations presented in the present paper is shown in Fig. 1.

It is expected that the critical phase resembles a Luttinger liquid~\cite{HAL80,HAL81} which can be described by a conformal field theory with central
charge $c=1$
similar to the massless phase of XXZ model in a homogeneous field~\cite{Alcaraz_1995}. However, while the latter model may be diagonalized using the Bethe Ansatz, the XXZ model in a staggered field is not integrable.
\begin{figure}
\unitlength1cm
\begin{picture}(18,4.5)(0,0)
 \put(1,0)  {\includegraphics[width=6.5cm]{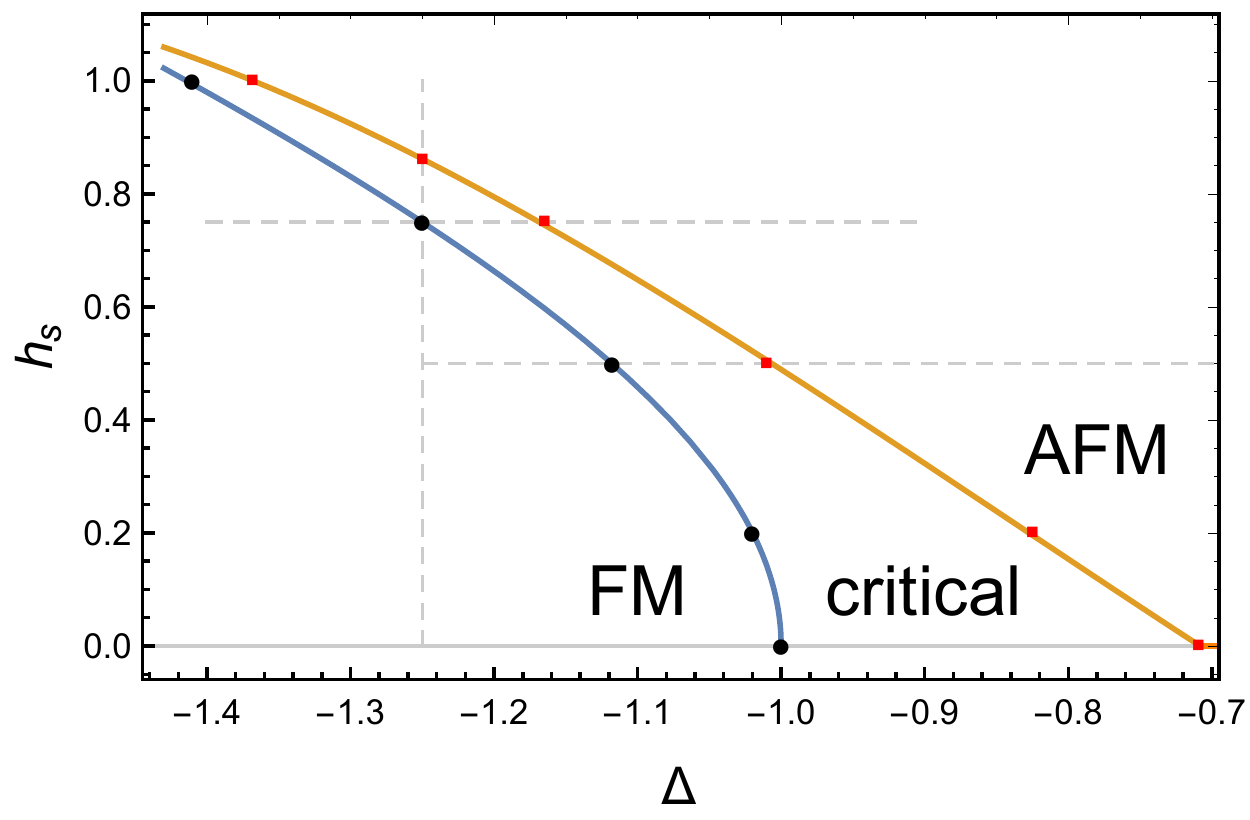}}
\end{picture}
\caption{\footnotesize Phase diagram  in the $(\Delta, h_s)$-plane for $h_s \ge 0$.
The blue line separates
the ferromagnetic phase (FM) from the critical phase, and the orange line separates the critical phase from the
anti-ferromagnetic phase (AFM). The light dashed lines indicate the cross sections where calculations are performed.
\label{fig:phaseDiagram}}
\end{figure}

Conformal symmetry relates the finite size spectrum of a discrete model to corresponding field theoretical operators, such that their
scaling dimensions may be determined from a numerical diagonalization of the model Hamiltonian.
Using various methods, notably the Bethe Ansatz, scaling dimensions were determined numerically
for the XXZ model (with and without homogeneous magnetic field)  and compared to field theoretical predictions~\cite{PhysRevLett.58.771, Woynarovich_1989}.
For the XXZ model in a staggered field such a comparison was attempted using exact diagonalization
for system sizes up to 18 sites~\cite{Alcaraz_1995, Okamoto_1996}.

Tensor renormalization is able to obtain spectra for much larger systems, and
in a recent paper~\cite{PhysRevB.100.134434} we studied the XXZ
chain in a longitudinal homogeneous field and showed by comparison to Bethe ansatz results that the method is able
to determine accurately the spectrum and phase diagram of such systems.
Here, we present a
detailed analysis of the XXZ model in a staggered field using the same method. We do this with two goals in mind: firstly, we determine the phase diagram
from the spectral gap and various other observables. Secondly, for the massless phase, we calculate  scaling dimensions from the finite size dependence of the spectrum.
These scaling dimensions are compared to the predictions of Luttinger liquid low energy field theory in an attempt to determine the validity of
the field theoretical description. In order to do so, the numerical uncertainties of the calculations must be assessed carefully.

The tensor renormalization method employed here is discussed in some detail
in our previous paper~\cite{PhysRevB.100.134434}. It is based on the HOTRG method
introduced in Ref.~\cite{XIE2012} with U(1) symmetry of the tensors implemented
explicitly. This enables large tensor sizes and labels the
calculated spectra with the appropriate U(1) quantum numbers. This is important for
the comparison with field theoretical results.
The low lying spectrum is determined from the calculated transfer matrix, which is directly
obtained from the coarse grained tensors. The renormalized tensor at each coarse graining step corresponds
to a certain system size, therefore, one obtains the complete finite size dependence of the spectrum required for the
determination of the scaling dimensions in a single run of the tensor renormalization procedure.

In section~\ref{sec-XXZ} we determine the phase diagram from the low lying spectrum. A detailed discussion of its
finite size dependence and a comparison with field theoretical predictions are presented in section~\ref{sec-finite-size}.
Details of our HOTRG tensor renormalization method are presented in section~\ref{sec-ten}. In particular, various
numerical issues that contribute to the overall uncertainty of the numerical results are addressed.
\begin{figure*}[t]
\unitlength1cm
\begin{picture}(18,4.)(0,0)
 \put(12,0)      {\includegraphics[width=5.cm]{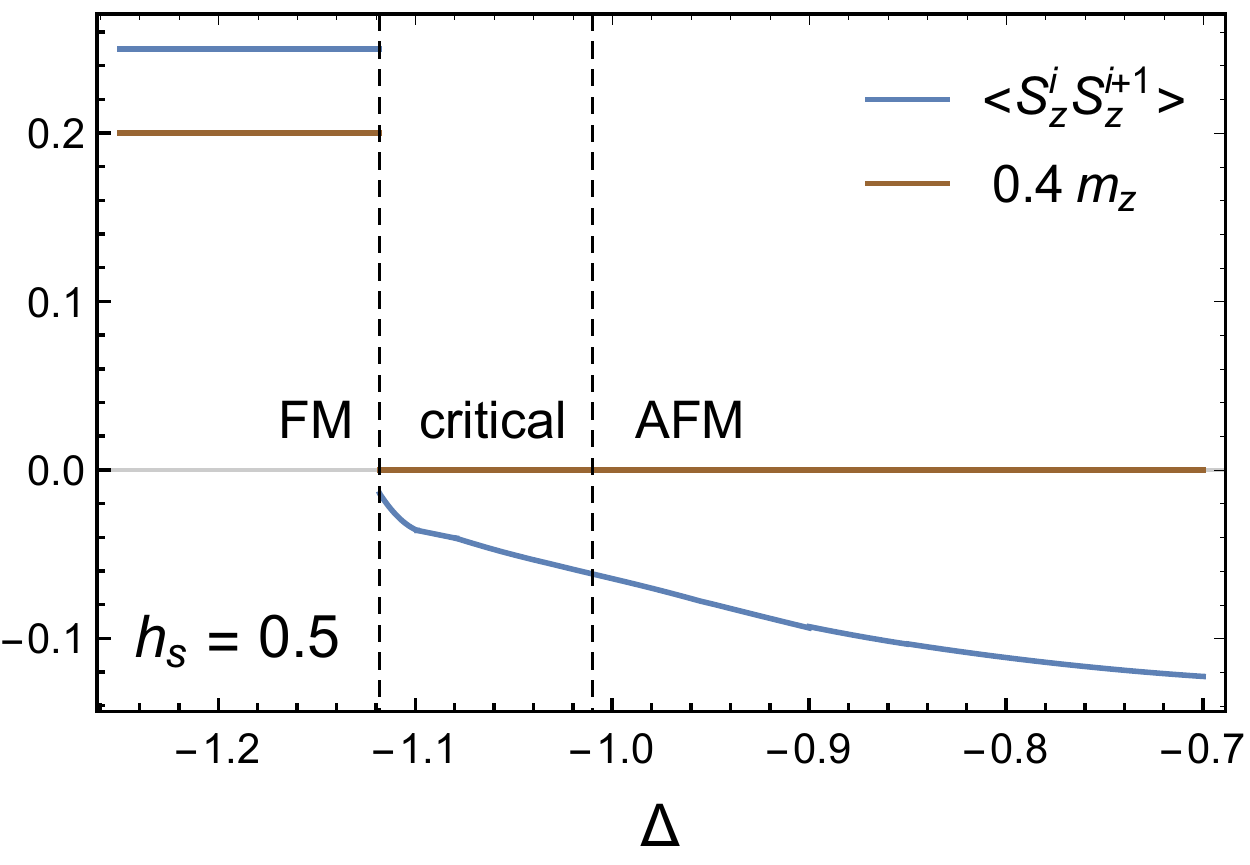}}
 \put(0,0)    {\includegraphics[width=5.cm]{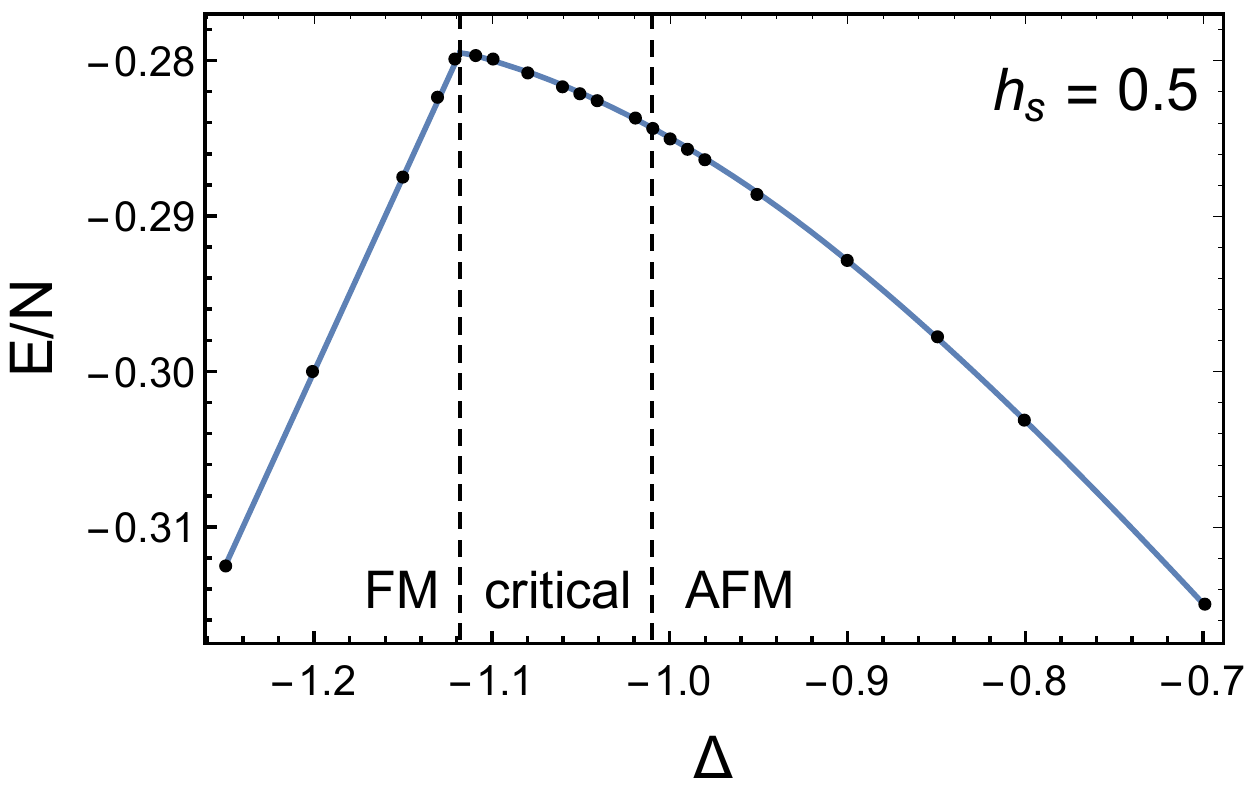}}
 \put(6,0)    {\includegraphics[width=5.cm]{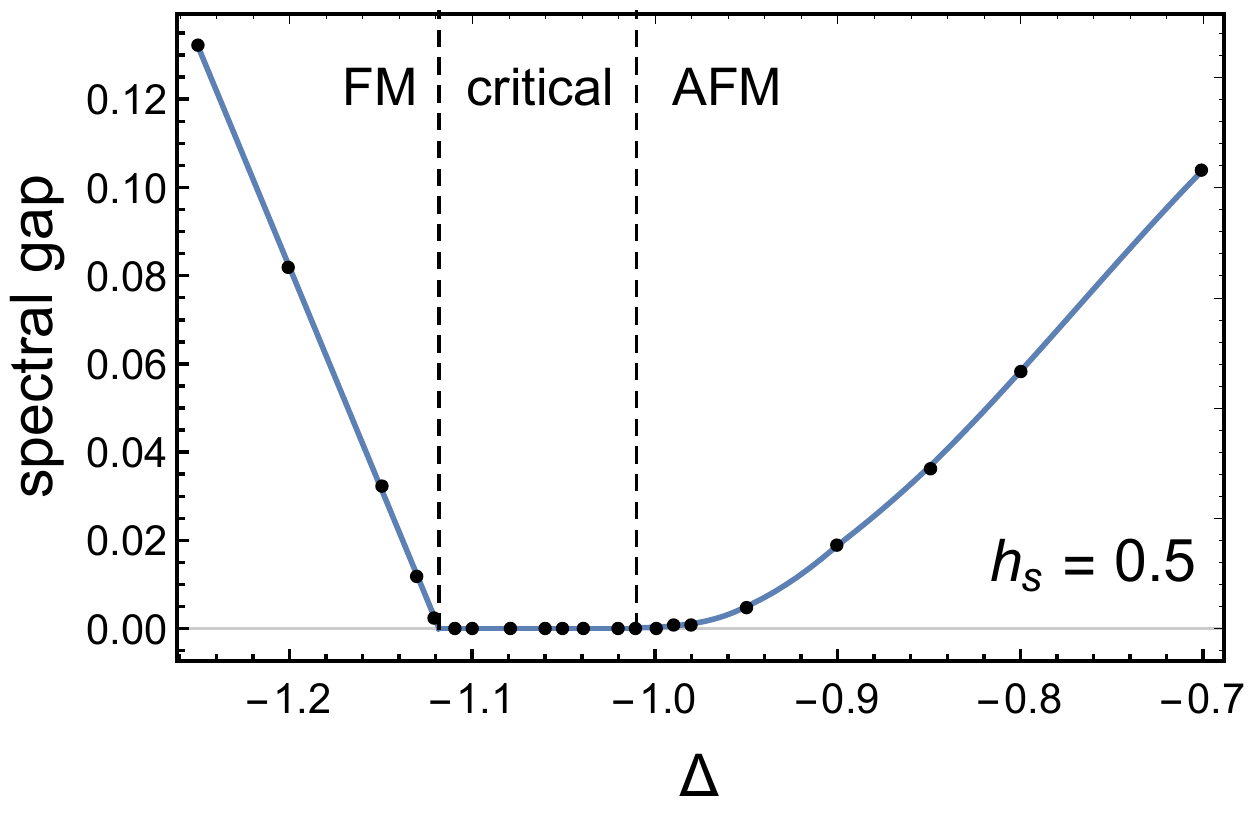}}
\end{picture}

\begin{picture}(18,4.)(0,0)
 \put(12,0)      {\includegraphics[width=5.cm]{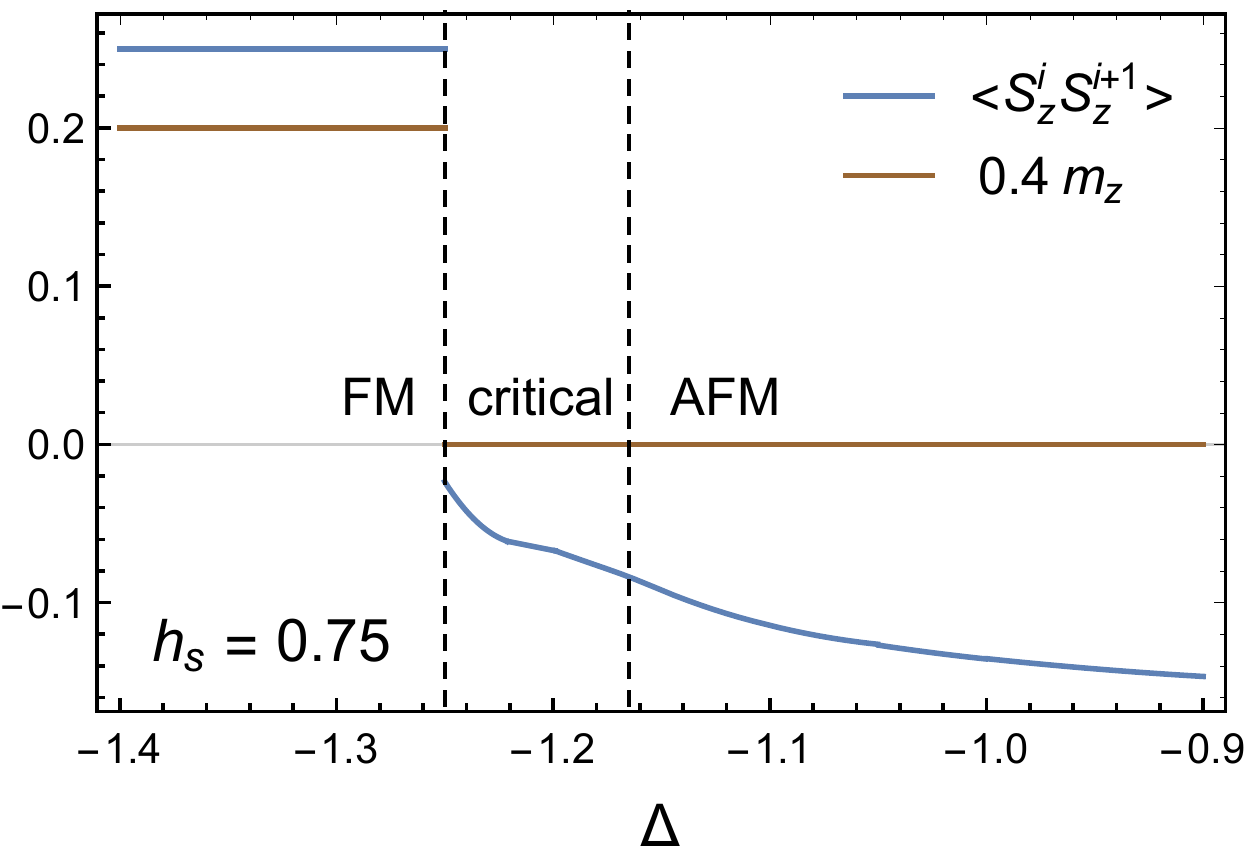}}
 \put(0,0)    {\includegraphics[width=5.cm]{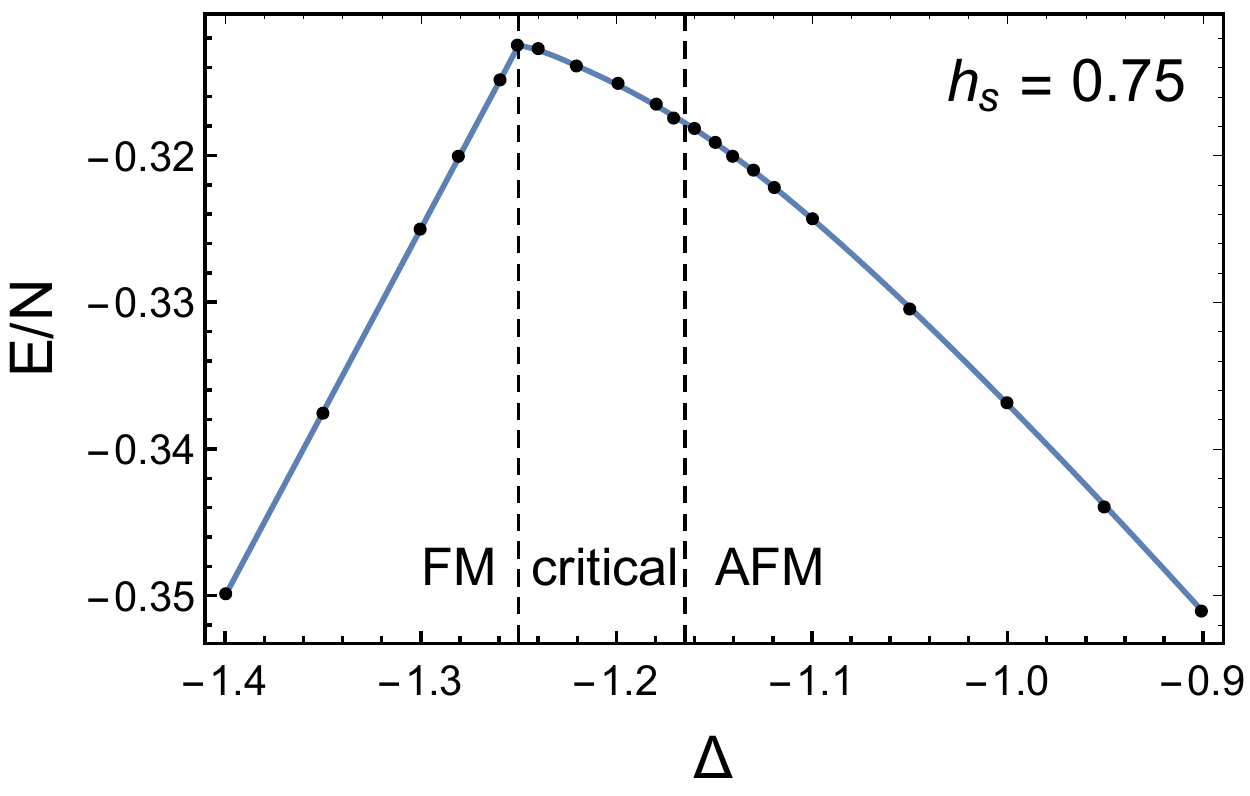}}
 \put(6,0)    {\includegraphics[width=5.cm]{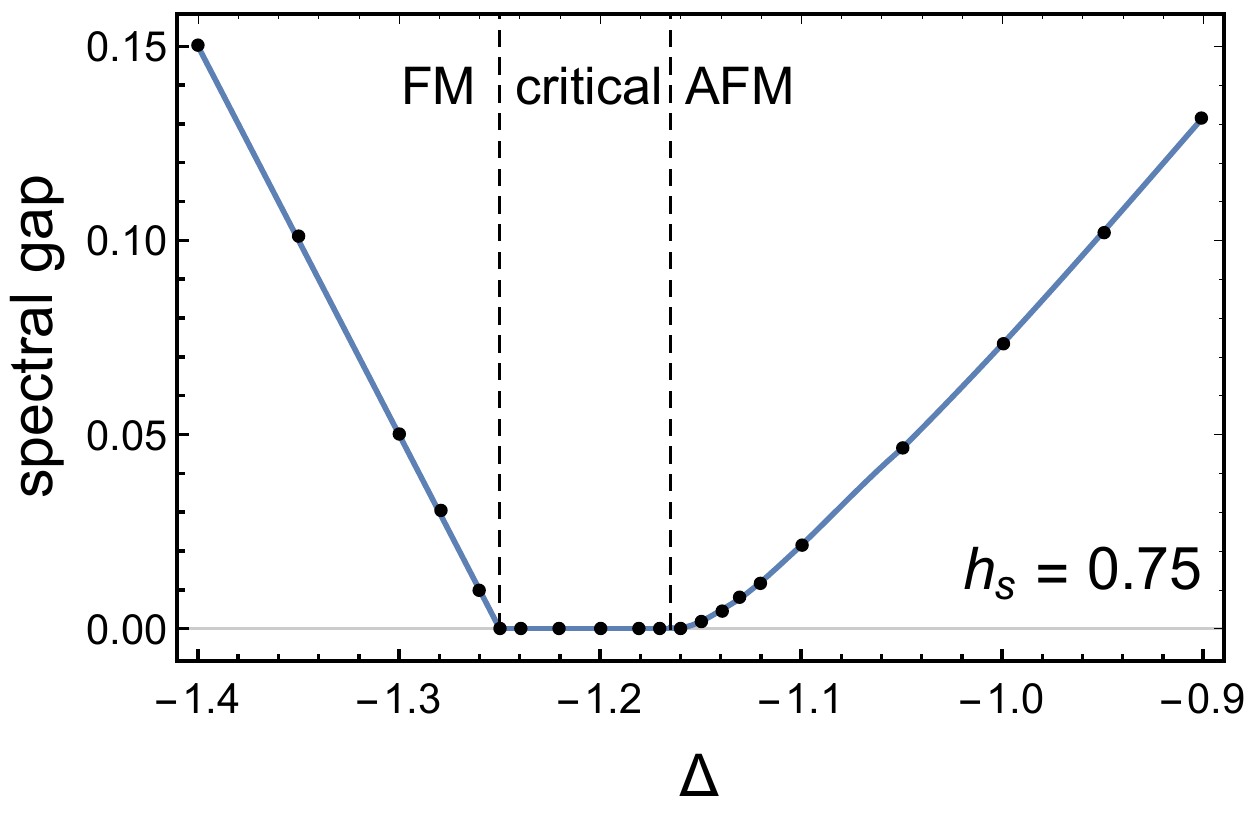}}
\end{picture}
\caption{\footnotesize Properties of the spin-1/2 XXZ model as a function of the anisotropy $\Delta$ at staggered magnetic fields $h_s=0.5$ and $h_s=0.75$.\\
(left) ground state energy per site.
(center) excitation gap.
(right) $z-z$ correlator $\langle S^i_z S_z^{i+1}\rangle$ and rescaled longitudinal magnetization $m_z$. \\
{The graphs for energy per site and spectral gap are qualitatively similar to Bethe ansatz results for $h_s=0$~\cite{PhysRevB.100.134434}. The magnetization in each phase is exactly the same as for $h_s=0$.} \\
The ground state energy shows a cusp at the FM phase boundary $\Delta_1 \simeq -1.118$ for $h_s=0.5$ and $\Delta_1 \simeq -1.250$ for $h_s=0.75$, respectively, in excellent agreement with the Pokrovsky-Talapov prediction $\Delta_1=-\sqrt{1+h_s^2}$. At $\Delta_1$
the gap closes, the magnetization jumps from 1/2 to 0, and the $z-z$ correlator changes discontinuously from 1/4 to $\simeq -0.014$ for $h_s=0.5$ and to $\simeq -0.024$ for $h_s=0.75$ {(in contrast to $h_s=0$, where it changes from 1/4 strictly to zero).} \\
At the AFM phase boundary $\Delta_2 \simeq -1.010$ for $h_s=0.5$  ($\Delta_2 \simeq -1.165$ for $h_s=0.75$) the spectral gap reopens smoothly, i.e., there is no cusp.
Neither the ground state energy nor the $z-z$ correlator or the magnetization show a discontinuity at that point.
\label{fig:XXZstagh075}}
\begin{picture}(18,4.)(0,0)
 \put(0,0)     {\includegraphics[width=5.cm]{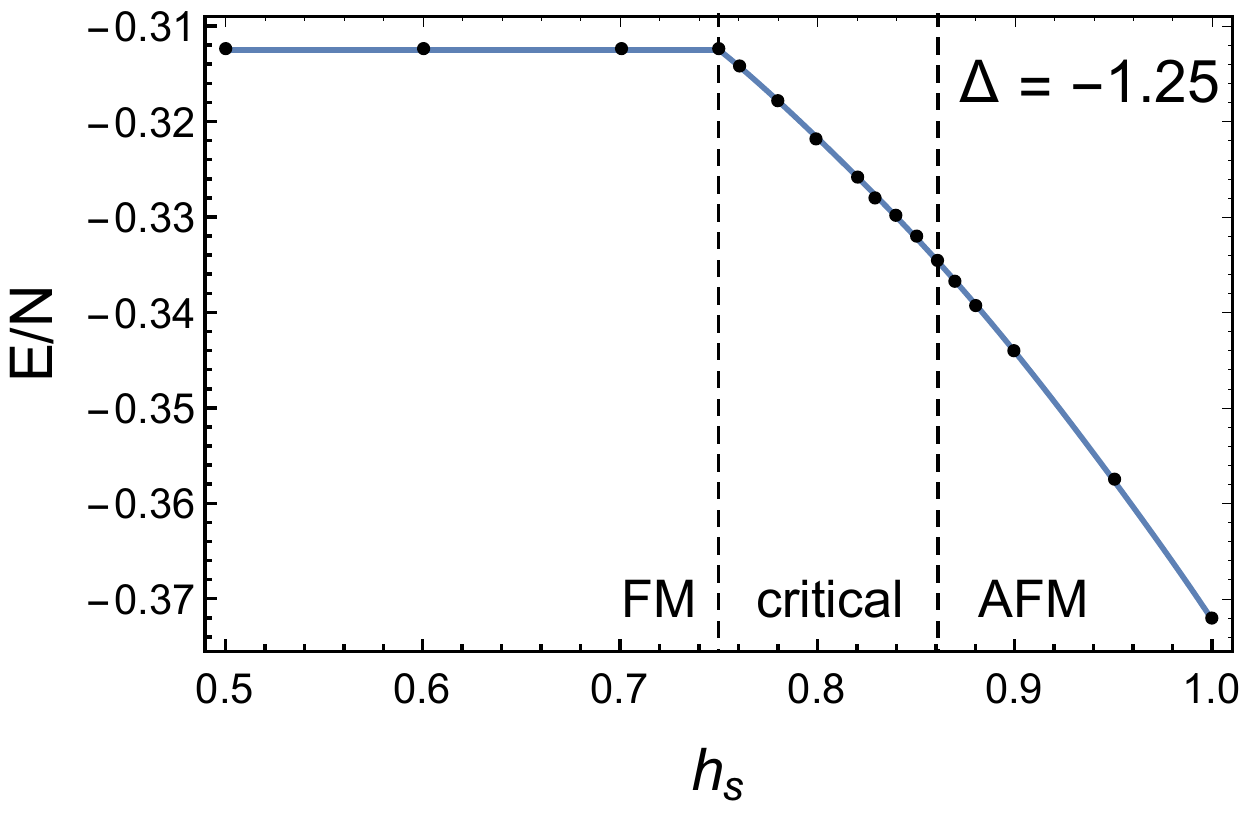}}
 \put(6,0)     {\includegraphics[width=5.cm]{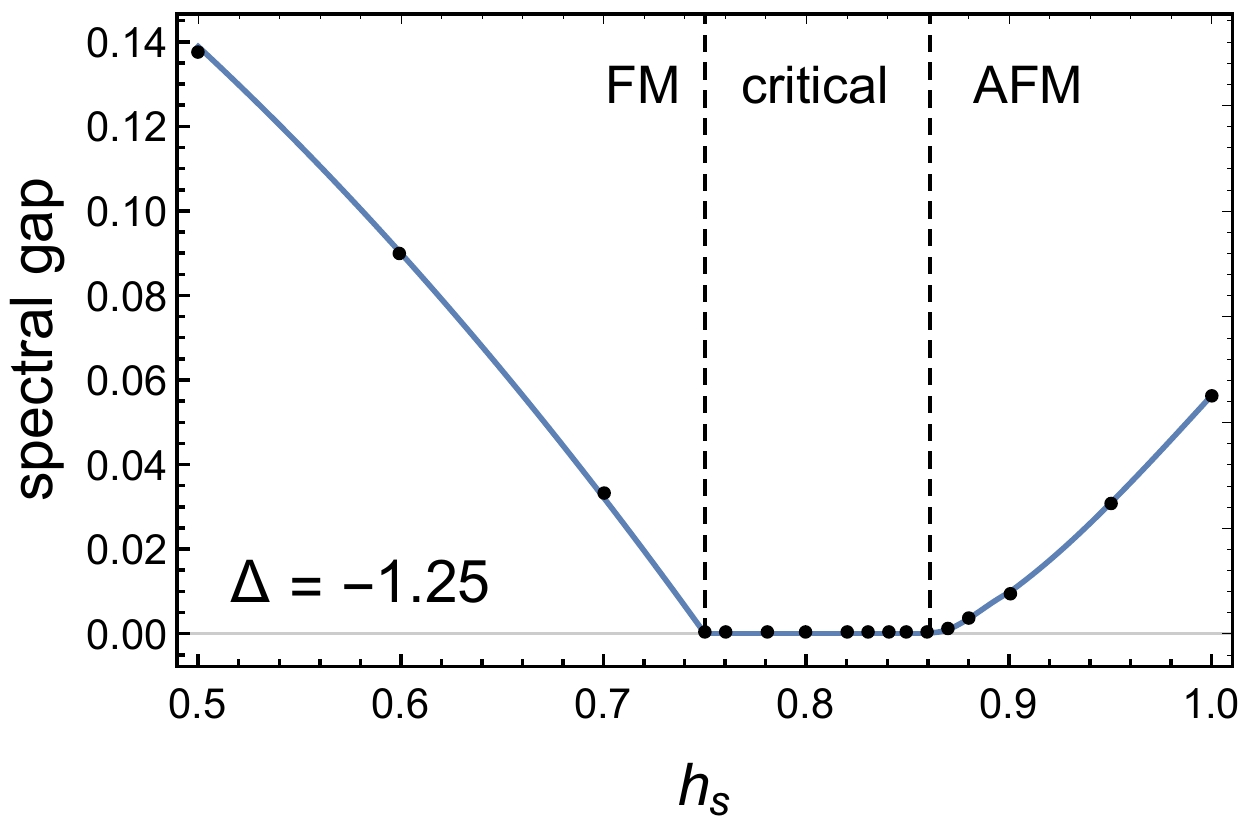}}
 \put(12,0)    {\includegraphics[width=5.cm]{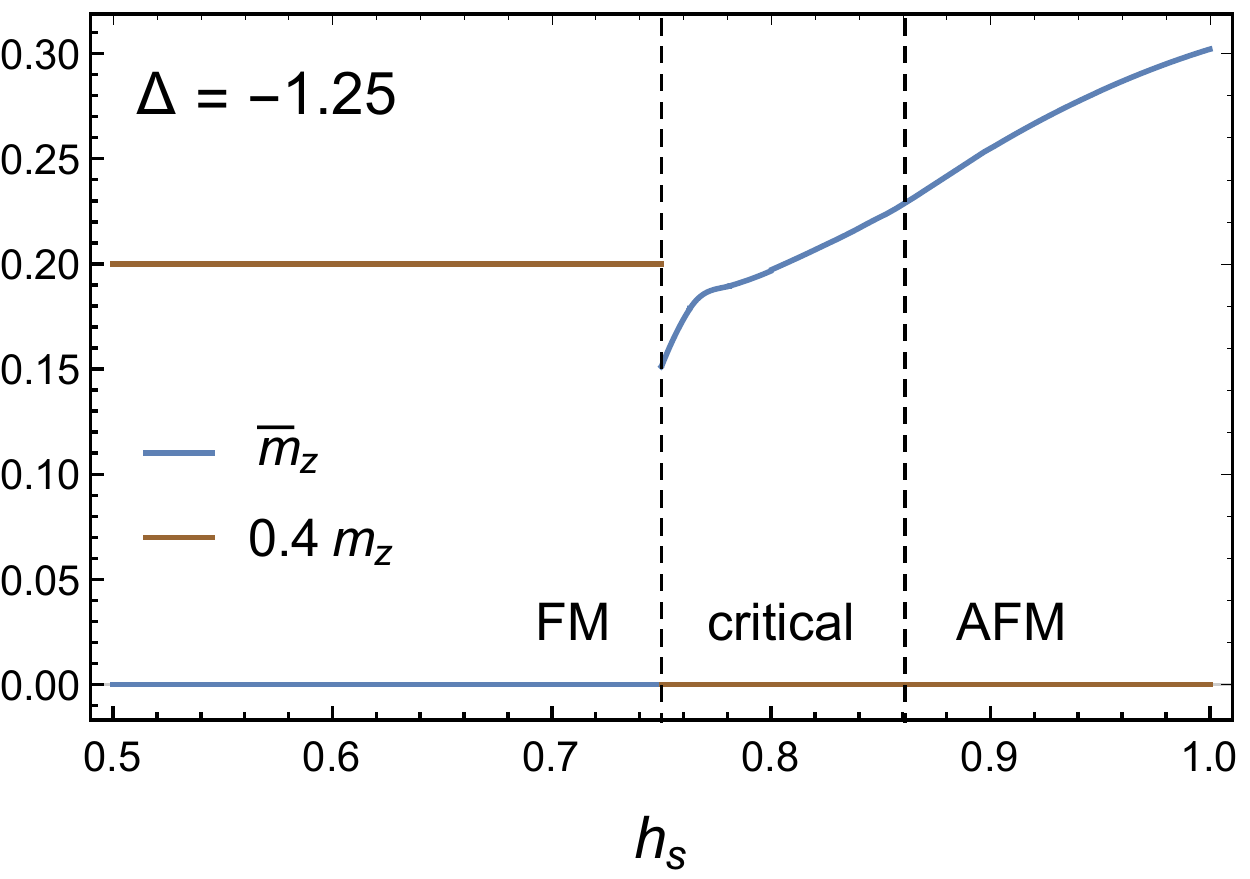}}
\end{picture}
\caption{\footnotesize  Properties of the spin-1/2 XXZ model as a function of the staggered magnetic field $h_s$ at $\Delta=-1.25$.\\
(left) ground state energy per site.
(center)  excitation gap.
(right) rescaled magnetization $m_z$ and staggered magnetization $\bar{m}_z$.\\
The energy is constant in the FM phase and shows a cusp at the critical point $h_{s1} \simeq 0.750$ in agreement with the prediction $h_{s1}=\sqrt{\Delta^2-1}$.\\
The energy gap is described precisely by the function $0.25 \, (1-(\frac{h_s}{0.75})^2)$ in the FM phase, and it shows a cusp at $h_{s1}$. At this point the gap closes,
the magnetization $m_z$ jumps from 1/2 to 0 while the staggered magnetization $\bar{m}_z$ jumps from 0 to $ \simeq 0.15$.\\
At $h_{s2} \simeq 0.861$ the gap smoothly reopens, i.e., without a cusp. Neither the ground state energy nor  the magnetizations show a discontinuity at that point.
\label{fig:gseXXZDeltam125}}
\end{figure*}

\section{XXZ model in a staggered field: ground state energy and spectral gap}\label{sec-XXZ}

The anisotropic spin-1/2 XXZ model in a staggered field is given by the Hamiltonian
\begin{equation}\label{XXZham}
H=\sum_{i=1}^N  S_x^i  S_x^{i+1}+ S_y^i  S_y^{i+1}+\Delta S_z^i  S_z^{i+1}-\ (-1)^i h_s S_z^i.
\end{equation}
Here, the $S^i_\lambda$ are spin-1/2 matrix representations of SU(2) and $N$ denotes the system size.
The model depends on two parameters: the anisotropy $\Delta$ and staggered field $h_s$.
We investigate its properties as a function of these two parameters.
The model is U(1) symmetric and its states may be labelled by the U(1) quantum numbers $S_z$ or by the magnetization $m_z={S_z}/{N}$.
The $Z_2$ spin-flip symmetry of the model is assumed to be broken by an infinitesimal longitudinal magnetic field,
since otherwise the magnetization in the ferromagnetic phase would vanish.

The U(1) symmetry is built into the numerical algorithm explicitly. The algorithm determines the spectrum of the Hamiltonian from a diagonalization
of the transfer matrix obtained from the HOTRG tensors as explained in more detail in section~\ref{sec-ten} and Ref.~\cite{PhysRevB.100.134434}.
Results for a number of selected parameters ($\Delta$, $h_s$) are shown in Figs.~\ref{fig:XXZstagh075} and~\ref{fig:gseXXZDeltam125}.
The parameters chosen for these figures correspond to the light dashed lines in the phase diagram (Fig.~\ref{fig:phaseDiagram}).
Strictly speaking, the presented numerical  results correspond to finite size systems of 1024 spins.
{ However, infinite-size results obtained by extrapolation of the finite-size calculations only differ within the resolution of the plots.}

The phase diagram of the model shown in Fig.~\ref{fig:phaseDiagram} is obtained from the calculations of the spectral properties, and the phase boundaries delineate
the parameter region where the gap is found to be numerically smaller than {$10^{-5}$}. This limit is chosen in view of the systematic uncertainties of our calculations  and the fact that we treat a large but finite system. A discussion of the expected numerical uncertainties is relegated to section~\ref{sec-ten}.

The phase diagram is symmetric with respect to the line $h_s=0$, and we only show the area $h_s \ge 0$.
The FM phase is separated from the massless phase by the Pokrovsky-Talapov
line $h_s=\sqrt{\Delta^2-1}$ for $\Delta \le -1$ determined from the smallest staggered field compatible with a ferromagnetic ground state, i.e., a state in the sector
$S_z=\pm N/2$.  In fact, the numerical calculations always find
the FM ground state  in the $S_z=\pm N/2$ sector.
The phase transition from the FM phase to the critical phase is first-order: The ground state energy and the gap show a cusp, the magnetization in $z$-direction jumps
from 1/2 to zero, the staggered magnetization jumps from zero to a finite value, and the $z-z$-correlator  $\langle S^i_z S_z^{i+1}\rangle$ from 1/4 to a finite value.
(These values are explicitly given in the figure captions for specific parameter sets.)

There is no analytical result for the boundary between the AFM and massless phases, except for $h_s$ close to zero, where field-theoretical considerations predict the boundary to be at $\Delta=-1/\sqrt{2}$. Our numerical result is in good agreement with this prediction, which will be further discussed in the following section. This AFM boundary was suggested to be of infinite order in Ref.~\cite{Okamoto_1996}, and
our results support this prediction as indicated by the smooth onset of the gap in the AFM region:  we do not find indications of discontinuities at this phase boundary in the ground state energy, magnetization or
$z-z$-correlator (as far as this is possible numerically).

\section{Finite size dependence in the massless phase}\label{sec-finite-size}

The energy spectrum as a function of the number of spins is immediately obtained
from the transfer matrices calculated from the HOTRG tensors at each `space' renormalization step.
As explained in more detail in the following section, the HOTRG coarse graining procedure involves time steps and space steps,
and at each space step the system size increases by a factor of two, starting from a system with  $N=2$.
Since we use a U(1) symmetric HOTRG implementation,
the transfer matrix calculated at each step is block diagonal. Each block
corresponds to a fixed number $r$ of `up' spins and is labelled by the spin-wave quantum number { $S_z=r-N/2$}.

\begin{figure*}
\unitlength1cm
\begin{picture}(18,4.5)(0,0)
 \put(5.7,0)      {\includegraphics[width=5.5cm]{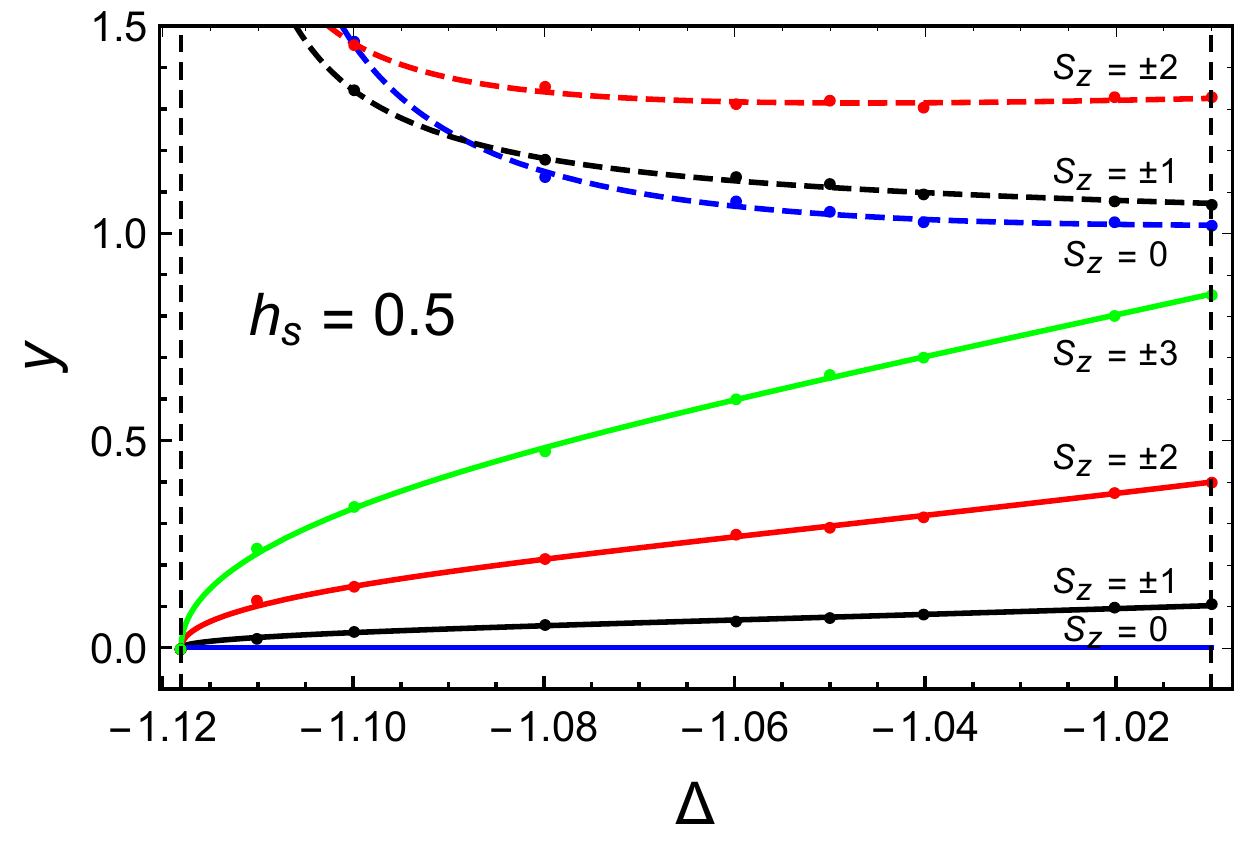}}
 \put(0,0)        {\includegraphics[width=5.5cm]{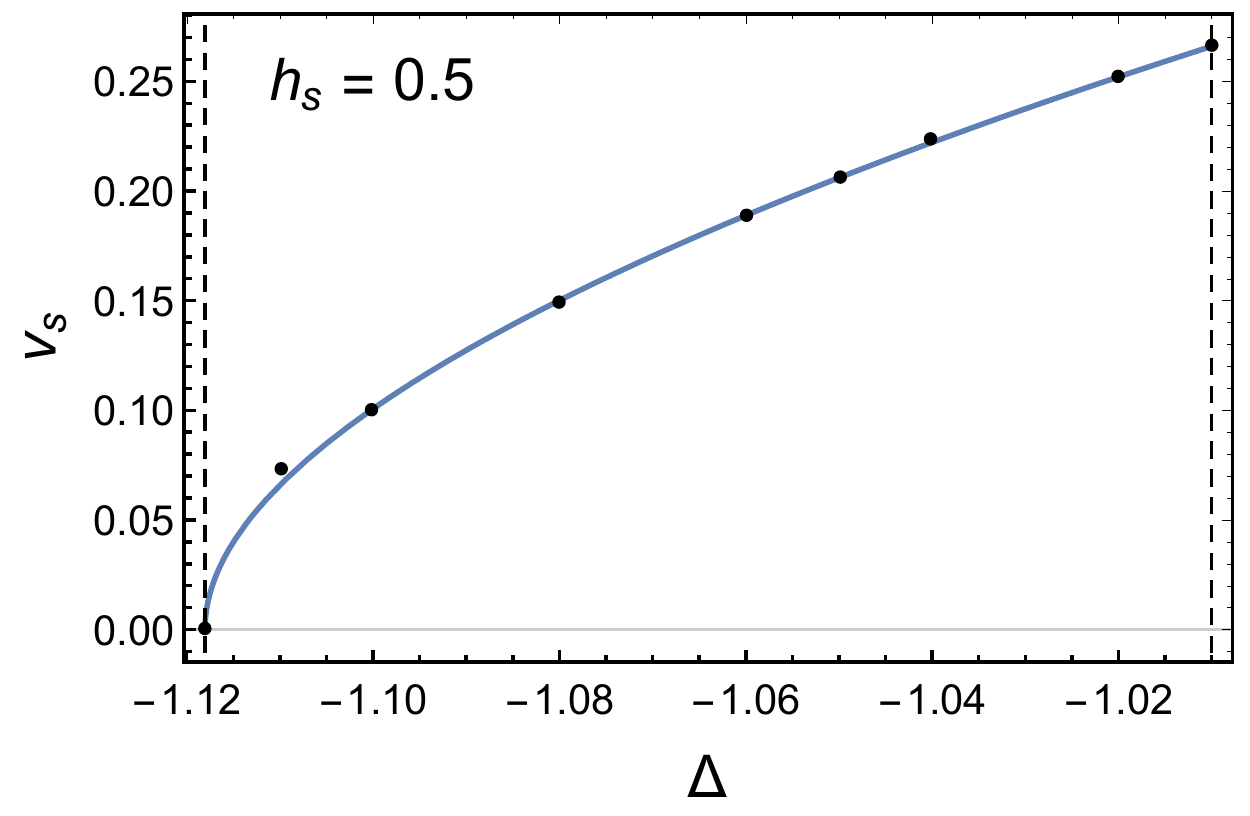}}
 \put(11.6,0)       {\includegraphics[width=5.2cm]{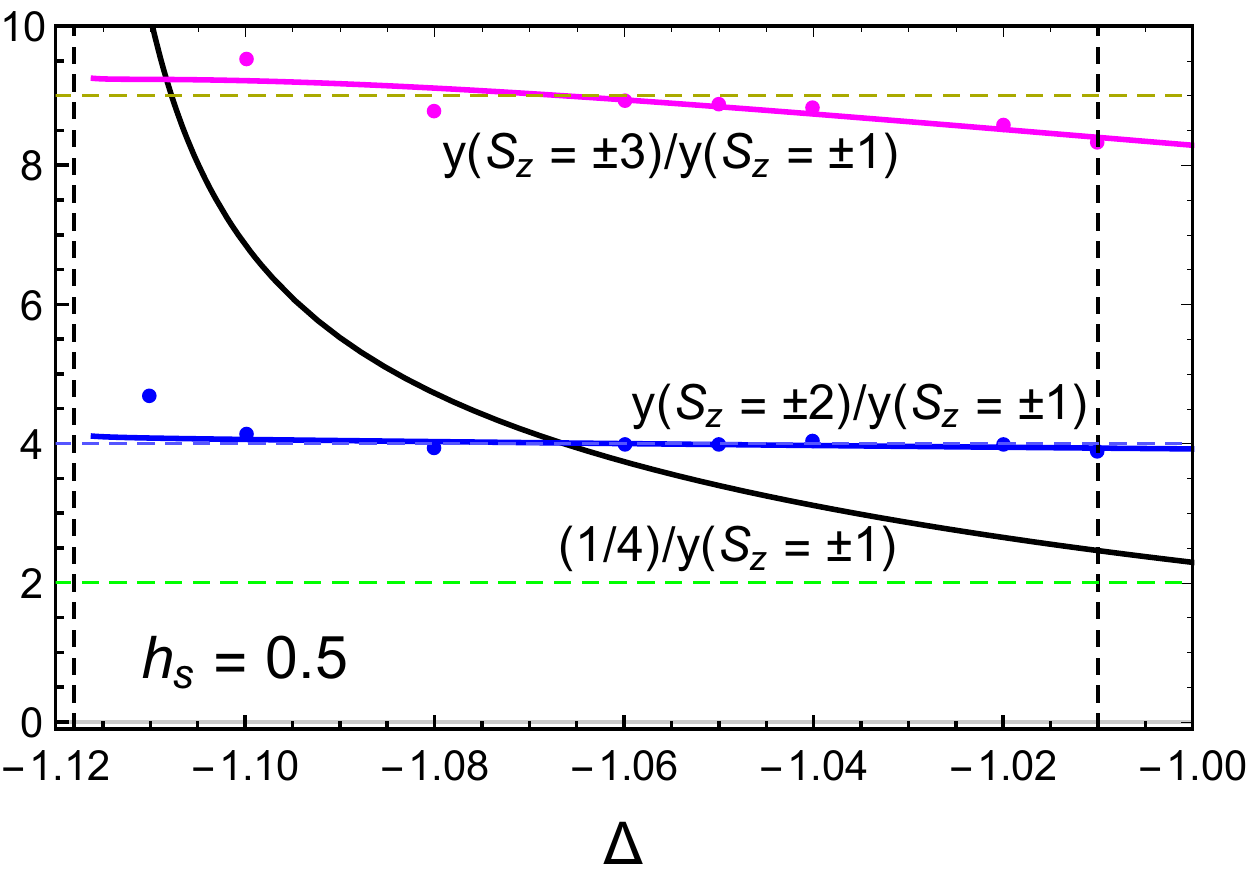}}
\end{picture}
\begin{picture}(18,4.5)(0,0)
 \put(5.7,0)      {\includegraphics[width=5.5cm]{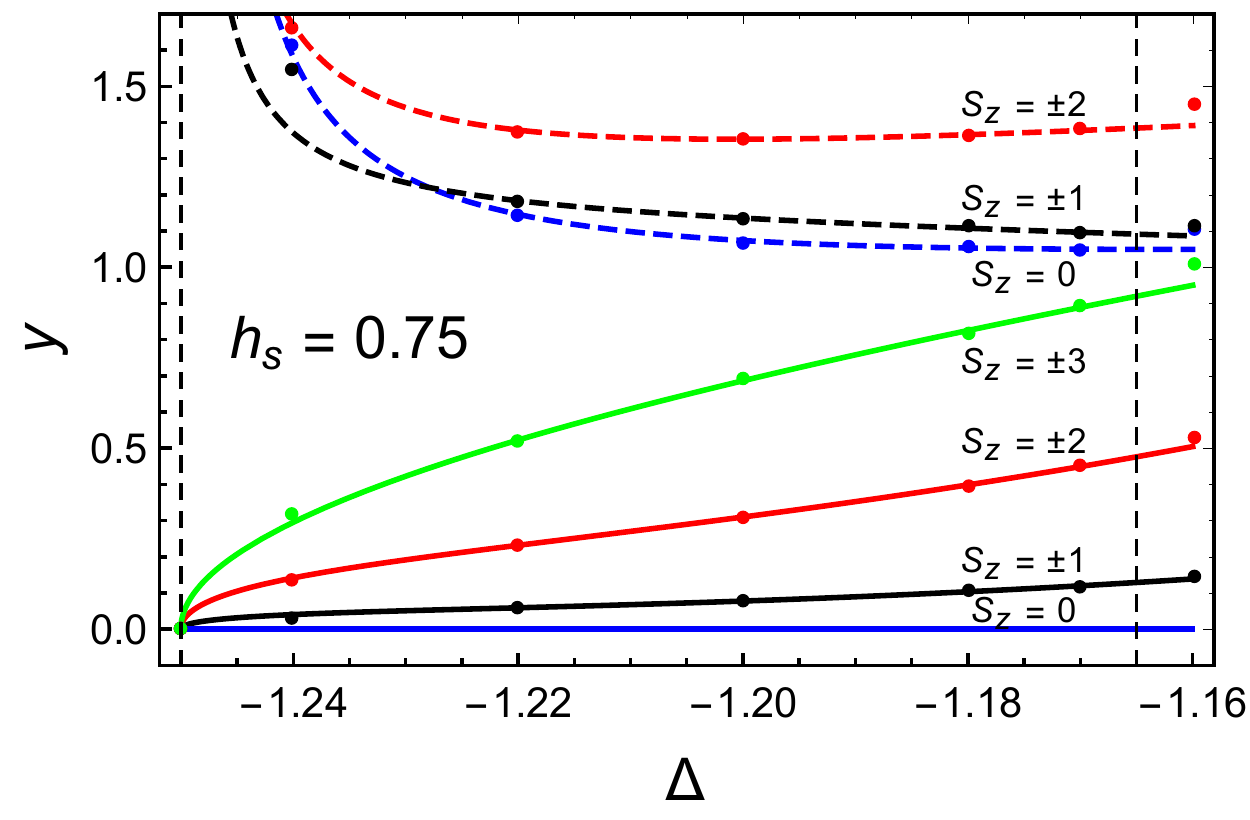}}
 \put(0,0)    {\includegraphics[width=5.5cm]{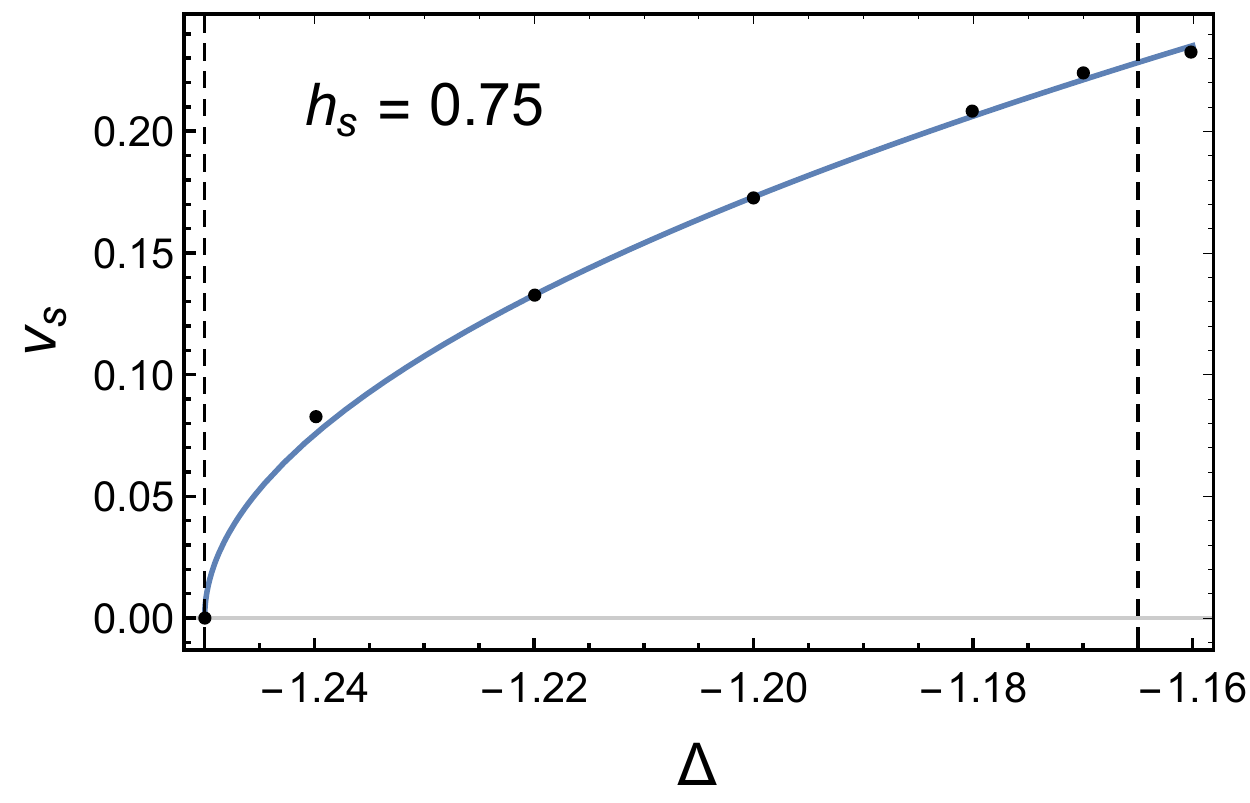}}
 \put(11.6,0)      {\includegraphics[width=5.2cm]{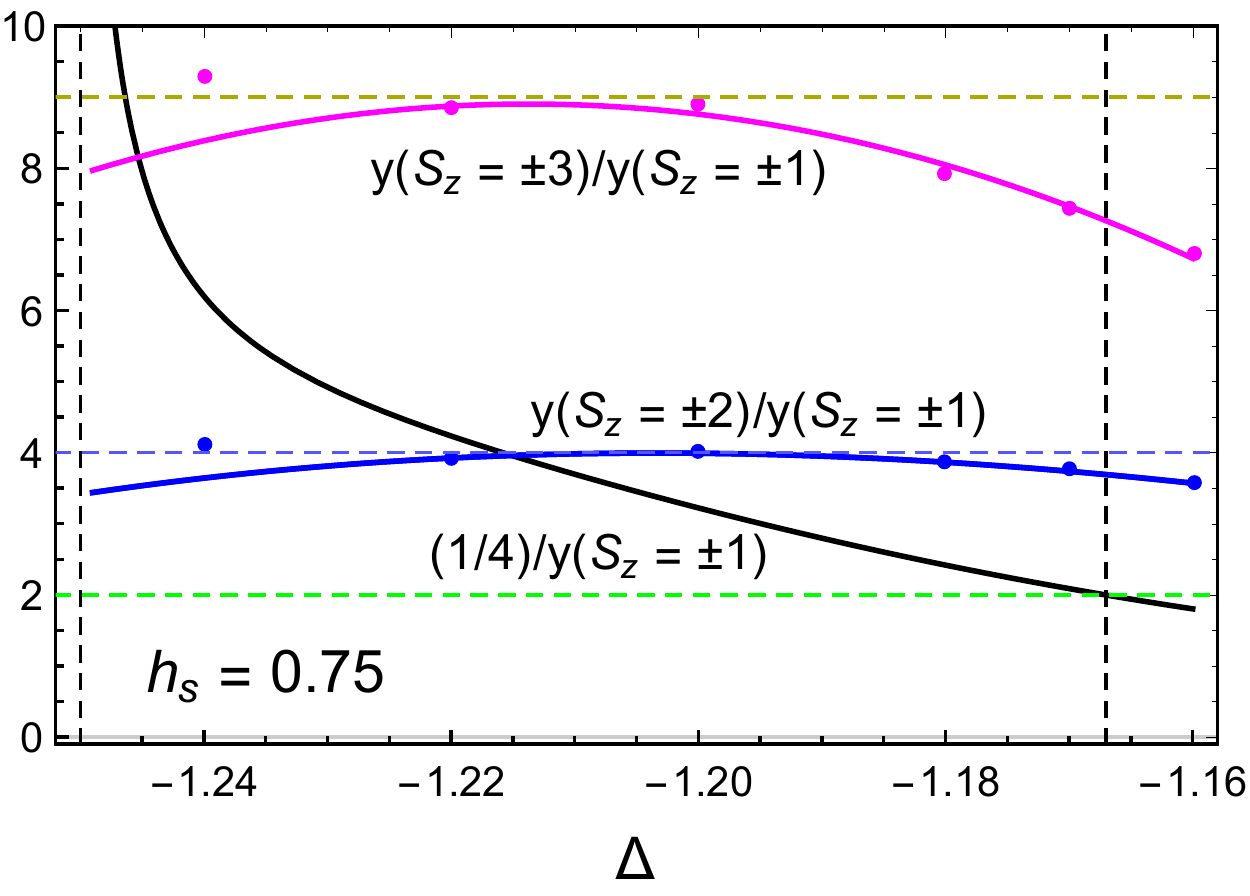}}
\end{picture}
\caption{\footnotesize Properties of the spin-1/2 XXZ model in the massless phase at $h_s=0.5$ and  $h_s=0.75$ as a function of the anisotropy $\Delta$.
The vertical dashed lines correspond to the phase boundaries as determined from the results in section~\ref{sec-XXZ}.
(left) spin wave velocity $v_s$.
(center) conformal towers of scaling dimensions $y$. The scaling dimension $y=0$ with $S_z=0$ corresponds to the non-degenerate ground state.
(right) ratios of scaling dimensions and $y=1/(4x_p)$. { Theoretical predictions for the ratios of the scaling dimensions for $S_z=\pm 2$ and $S_z=\pm 3$ to the scaling dimension for $S_z=\pm 1$ are shown by blue and orange dashed lines, respectively. The crossing of the quantity $y=1/(4x_p)$ with the green dashed line corresponds to a sign change of the scaling dimension of the perturbing operator $y_p=2-1/{(4 x_p)}$, which indicates the massless-to-AFM phase transition.}
\label{fig:XXZstagh075scaling}}
\begin{picture}(18,4.5)(0,0)
 \put(5.7,0)   {\includegraphics[width=5.5cm]{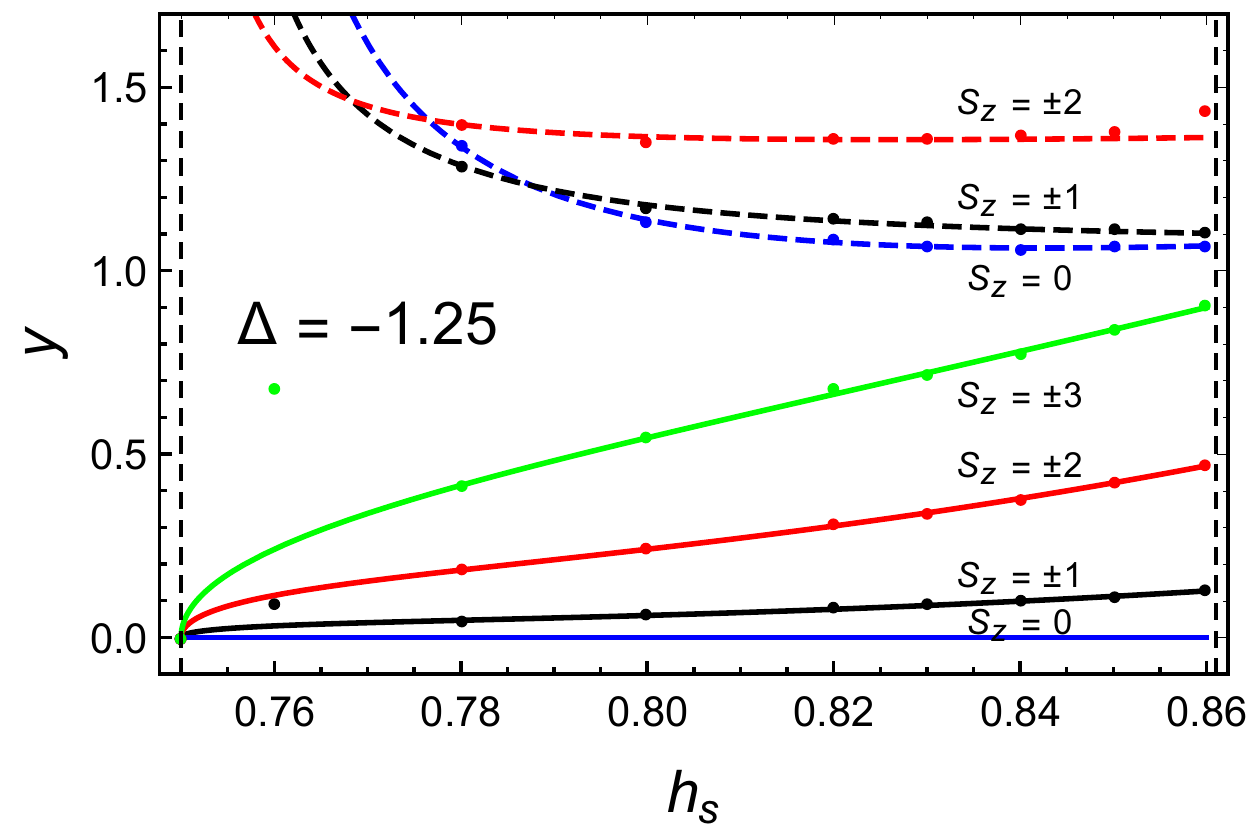}}
 \put(0,0)    {\includegraphics[width=5.5cm]{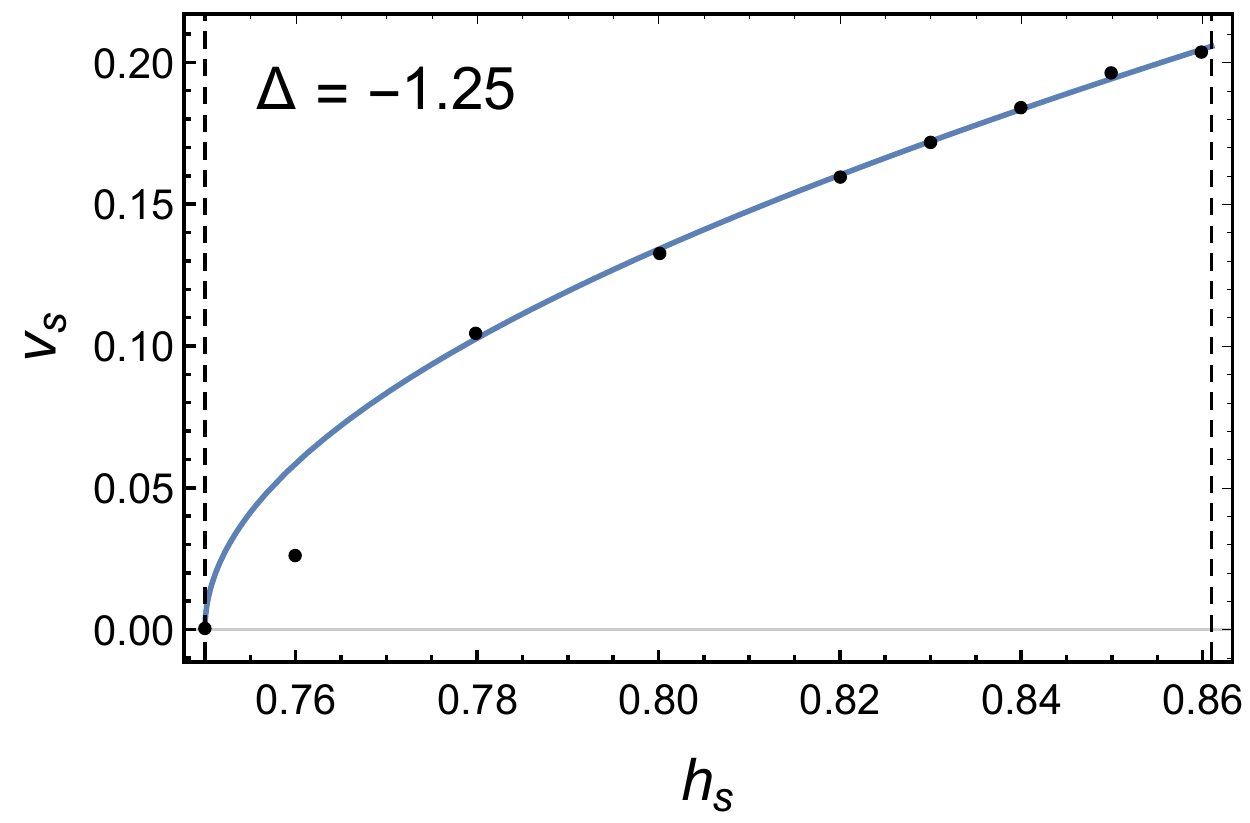}}
 \put(11.6,0)    {\includegraphics[width=5.2cm]{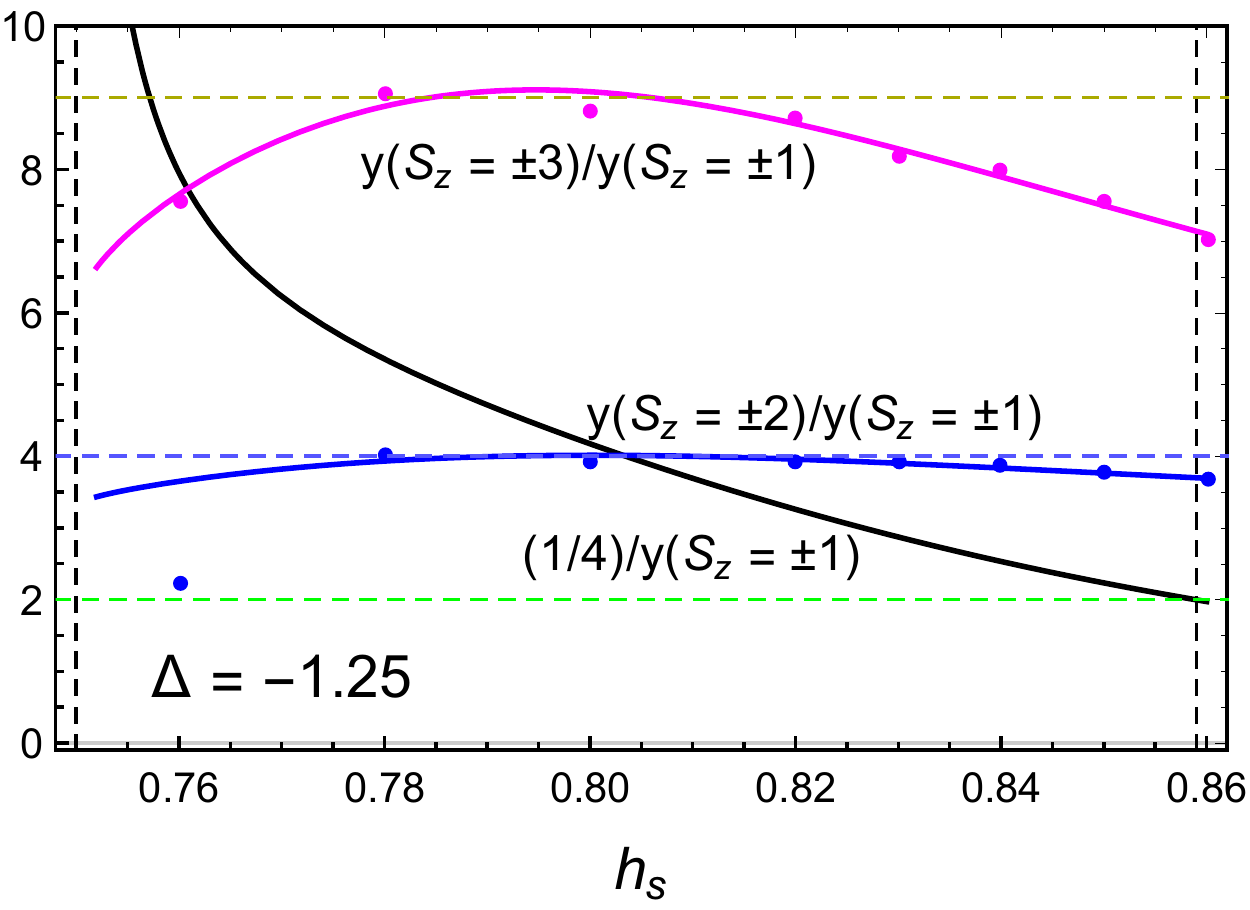}}
\end{picture}
\caption{\footnotesize  Properties of the spin-1/2 XXZ model in the massless phase at $\Delta=-1.25$ as a function of the staggered field $h_s$. Details as in Fig.~\ref{fig:XXZstagh075scaling} above.
\label{fig:gseXXZDeltam125scaling}}
\end{figure*}

For the size dependence $E_0(N)$ of the ground state energy in the massless regime, conformal invariance predicts~\cite{CAR86}
\begin{equation}\label{finite-size-gse}
\frac{E_0(N)}{N}=\epsilon-\frac{\pi c v_s}{6 N^2} +o(N^{-2})
\end{equation}
with $\epsilon$ the ground state energy per site of the infinite system, $c$ the central charge and $v_s$ the spin-wave velocity.
Assuming that $c=1$, we determine $v_s$ from numerical data for the ground state energy.
This (absolute) ground state in the massless phase is found in the $S_z=0$ sector.
Selected results for the spin-wave velocities are shown in
Fig.~\ref{fig:XXZstagh075scaling} for the staggered fields $h_s=0.5$ and $h_s=0.75$, respectively.
Results for $\Delta=-1.25$ as a function of $h_s$ are shown in Fig.~\ref{fig:gseXXZDeltam125scaling}.
As already mentioned, these particular parameter selections  are indicated in the
phase diagram (Fig.~\ref{fig:phaseDiagram}) by the light grey lines. In addition to the numerical data, we show
fits to these results. These fits assume a square root singularity of $v_s$ at the FM phase boundary. Such a dependence is suggested by
the Bethe Ansatz result for zero staggered field, but for finite fields its is a conjecture, however supported by the numerical
data. Close to the FM boundary our numerical results show significant deviation from this fit, and in fact the renormalization
procedure does have difficulties to converge in this parameter region.

\begin{figure}
\unitlength1cm
\begin{picture}(8,4.)(0,0)
 \put(1,0)  {\includegraphics[width=6cm]{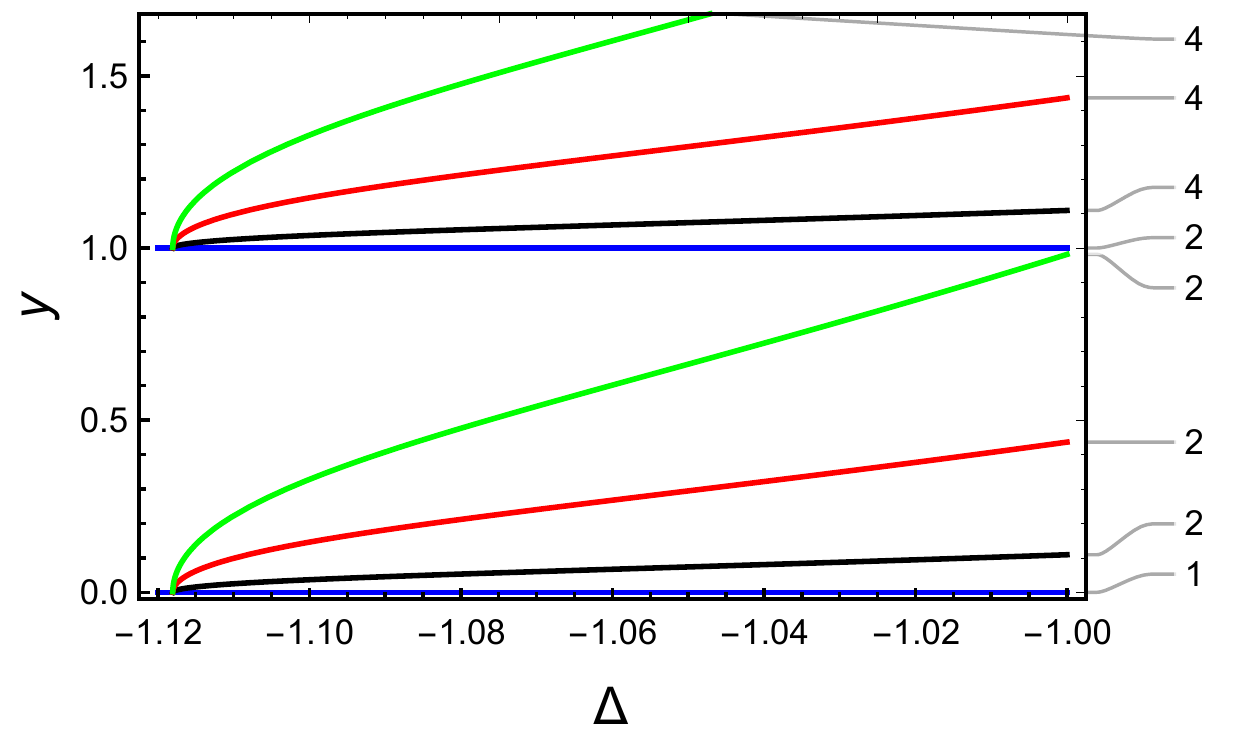}}
\end{picture}
\caption{\footnotesize   Prediction of Eq.~(\ref{luttliq}) for $h_s=0.5$ using the same color scheme as in Fig.~\ref{fig:XXZstagh075scaling} with  $x_p$  determined from the
tensor network data; the numbers to the right indicate the degeneracy of the states. The degeneracies arise due to the $j$ and $j^\prime$ quantum numbers in Eq.~(\ref{luttliq}),
which take only the values 0 and 1 here. As a consequence, the higher spectrum is just a shift of the lower spectrum by one unit with an increase of degeneracy.
\label{fig:lutt}}
\end{figure}

For the size dependence of the gaps, conformal invariance predicts~\cite{CAR86}
\begin{equation}\label{finite-size-gaps}
E_{\alpha}(N)-E_0(N)= \frac{2\pi v_s}{N} y_\alpha  +o(N^{-1})
\end{equation}
which determines the scaling dimensions $y_\alpha$ using the spin-wave velocities obtained from the ground state energy as input. In
Figs.~\ref{fig:XXZstagh075scaling} and \ref{fig:gseXXZDeltam125scaling} we show the scaling dimensions corresponding to
the ground state and first excited states in the $S_z=\pm 0$, $S_z=\pm 1$, $S_z=\pm 2$, and  $S_z=\pm 3$  sectors. The (approximate) degeneracies of these states
are found to be compatible with those indicated in Fig.~\ref{fig:lutt}, i.e., the results shown in the Figs.~\ref{fig:XXZstagh075scaling} and \ref{fig:gseXXZDeltam125scaling} correspond to 17 different
states including the (absolute) ground state, which are determined  from the diagonalization of the transfer matrix.
We observe that the scaling dimensions depend continuously on both $\Delta$ and $h_s$. For the lower part of the conformal tower  we show fits to our numerical data assuming
an arccosine singularity at the FM boundary, which is suggested by the Bethe Ansatz results for zero staggered field.
The upper part is fitted assuming a hyperbolic singularity. This singularity is phenomenological and in disagreement with the field theoretical prediction as will
be further discussed below.

In order that the whole massless phase of the XXZ model in a staggered magnetic field
may be regarded a Luttinger liquid~\cite{PhysRevLett.58.771, Alcaraz_1995},
 it should be possible to express the scaling dimensions as
\begin{equation}\label{luttliq}
y_\alpha=x_{n,m}+j+j^\prime, \;\;\;\;\;x_{n,m}=n^2 x_p +  \frac{m^2}{4 x_p}.
\end{equation}
and non-negative integers $n$, $m$, $j$, and $j^\prime$. Here, $n=S_z$ while $j, j^\prime$ are spin-wave excitation numbers and $m$ is the vorticity.
Like the spin-wave velocity, the parameter $x_p=x_{1,0}$ is a smooth function of the model parameters $\Delta$ and $h_s$.  For $h_s=0$ this function can be easily determined from
Bethe Ansatz, however, here it must be determined numerically from the lowest state in the $S_z=1$ sector. With this $x_p$ we predict the conformal tower
$y_\alpha$ from our data as shown in Fig.~\ref{fig:lutt}. Obviously, our numerical data are in reasonable agreement with the field theoretical prediction with the exception of the hyperbolic singularity seen close to the FM boundary for the upper part of the conformal tower. Specifically, the excited states in the various spin sectors are shifted by about one unit  $x_\alpha \sim 1$ suggesting that these states correspond to $j+j^\prime=1$ and $x_{0,0}=0$.
Clearly, the upper part of the  numerically determined conformal tower shows qualitative differences to the field theoretical result.
However, close to the FM boundary precise calculations are rather difficult, and more numerical data are needed to confirm this result.

A more quantitative comparison between Eq.~(\ref{luttliq}) and our data is attempted in the right column of Fig.~\ref{fig:XXZstagh075scaling} and \ref{fig:gseXXZDeltam125scaling}. In particular, we
check the predictions $y(S_z=\pm 2)/y(S_z=\pm 1)=4$ and $y(S_z=\pm 3)/y(S_z=\pm 1)=9$ for the lower part of the conformal tower. Here we see that this is quantitatively reproduced only in the
center of the massless phase. Close to the AFM phase boundary there are convergence problems of the renormalization procedure, but seemingly less severe than at the FM boundary.

We note that the scaling dimension $y=x_{0,1}=1/{(4x_p)}$ lies well above those shown in our figures, and cannot be easily identified
in the calculated spectra. However, we can infer this scaling dimension from the numerically determined $x_p$ as was done for Fig.~\ref{fig:lutt}.  This scaling dimension is of interest because it is related to  the scaling dimension  of
the perturbing operator which opens the gap at the AFM boundary of the massless phase: $y_p=2-1/{(4 x_p)}$~\cite{PhysRevLett.58.771, Alcaraz_1995}. Specifically, for $h_s=0$
one finds $x_p=\arccos(-\Delta)/2\pi$ using the Bethe Ansatz. As a consequence, the perturbing operator $y_p$ changes sign at $\Delta=-1/\sqrt{2}$, and according to renormalization group theory the
perturbing operator is relevant for $\Delta<-1/\sqrt{2}$, and a spectral gap opens at this point. This point marks the AFM phase boundary for $h_s=0$, which is reproduced by our numerical data
as discussed in the previous section. From the data for finite staggered fields shown in Figs.~\ref{fig:XXZstagh075scaling} and~\ref{fig:gseXXZDeltam125scaling} we see that $1/{(4 x_p)}$ passes through $y=2$ always at or close to the numerically determined phase boundary.
This suggests that scaling dimension $y_p=2-1/{(4 x_p)}$ of the perturbing operator for finite staggered fields is correctly predicted by the numerical data, and $y_p$ changes sign at the AFM phase boundary.
In conclusion, an interpretation  in terms of the
Luttinger liquid theory formulated by  Eq.~(\ref{luttliq}) is supported by the numerical data, but shows quantitative differences
close to the FM boundary.

\section{Tensor network implementation}\label{sec-ten}

Details of our U(1) symmetric tensor network implementation have been reported in Ref.~\cite{PhysRevB.100.134434},
therefore we can be brief here and only review a few key features and minor changes. However, we exhibit
various sources of error inherent in our method: truncation error, Trotter error, finite size and finite
temperature error.

\begin{figure*}
\unitlength1cm
\begin{picture}(18,5.)(0,0)
 \put(1,0)  {\includegraphics[width=16cm]{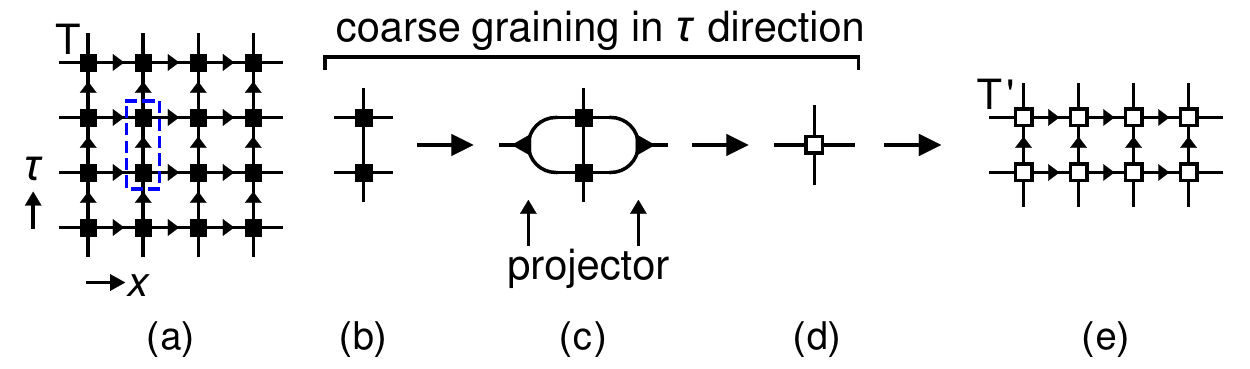}}
\end{picture}
\caption{\footnotesize   The essentials of the HOTRG coarse graining method: (a) a symmetric tensor network as a directed graph. (b) two tensors $T$ symbolized by the small black squares
are contracted into one. (c) unitary projection and approximation using HOTRG.
(d) the renormalized tensor. (e) the coarse grained tensor network.  (Arrows are omitted in (b), (c), (d) for simplicity.)
\label{fig:HOTRG}}
\end{figure*}

The partition function of a (1+1)D quantum system is expressed
as a tensor trace ~\cite{LEV2007, GU2009}
\begin{equation}\label{part func}
Z= {\rm Tr~} e^ {-\beta H}={\rm tTr~}  T^{\otimes K}.
\end{equation}
Here, $H$ is the Hamiltonian of the many-body system  and $\beta$ the inverse temperature. The (imaginary) time dimension
is discretized into $\tau=\beta/M$ time intervals, and there are $N$ spins in space direction.
The four-index tensors $T_{ijkl}$ obtained from the Hamiltonian depend on $\beta$ explicitly.
The tensor trace is determined
using the HOTRG coarse graining procedure as outlined in Ref.~\cite{XIE2012} and briefly summarized in
Fig.~\ref{fig:HOTRG}. The crucial step implementing the coarse graining is shown in Fig.~\ref{fig:HOTRG}(c). Here, the tensor size is reduced according to a higher order singular value decomposition~\cite{LAT06}, and only the dominant singular values are kept.
The main specific detail of our implementation as described in Ref.~\cite{PhysRevB.100.134434} is the explicit implementation of U(1) symmetry according
to the tensor structure proposed by Singh and Vidal~\cite{PhysRevA.82.050301, SIN11}. This not only saves
computing resources but assigns U(1) quantum numbers to calculated physical quantities.

At each coarse graining step either the inverse temperature or the space dimension increase by factor of two, while
the tensor size remains fixed above a certain system size, and the information about the system is `squeezed' into the `renormalized
tensors'. The finite size spectrum discussed in section~\ref{sec-finite-size} is obtained from these tensors at each space step. Since
the trace is not calculated exactly but via a coarse graining procedure, numerical uncertainties due to truncation are introduced.
These uncertainties are measured at each step from the {total `weight' of} neglected singular values.

The low lying spectrum is obtained from the tensor $T$ at every system size by diagonalization
of the transfer matrix
\begin{equation}
M_{ud}={\sum_l T_{luld}}
\end{equation}
where {$l$} corresponds to the space indices.
The transfer matrix is block diagonal due to the explicit implementation of U(1) symmetry, and each block is labelled by the appropriate U(1) quantum number.
These blocks are diagonalized, and the spectrum is obtained from the logarithm of the eigenvalues.
Other properties of the system, e.g., the staggered magnetization, can be determined from the parameter dependence of the spectrum.
The precision of the obtained spectrum depends on the truncation error, which in turn is related to the size the tensors we can handle.
The numerical results correspond to finite size systems of 1024 spins. However, infinite-size results differ only on the order of $~10^{-6}$.
Typically, the calculations are made with nominal tensor sizes of $m=130$ in each tensor dimension, not counting savings due to U(1) symmetry.


{ Unfortunately, the truncation error is not the only error to be considered. Two other errors are incurred: a) A Trotter error, which is due to the construction of the imaginary time discretization; b) the error caused by a large but finite $\beta$ (i.e. finite temperature). The Trotter error can obviously be minimized by choosing a smaller time step $\tau$ (we choose $\tau=0.001$ found by numerical experiments). However, then one faces the problem that such $\tau$ might be too small to reach a large enough $\beta$ within a limited number of time steps.

In view of this, for the calculations presented in this paper we changed the computer program such that we can do more imaginary time steps than space steps. First, at the start of the calculation, we do about 8 initial imaginary time steps without intermediate space steps. This modified `initial' tensor is truncated to a suitable size as to minimize the truncation error.
Empirically we found that for $m=16$ the discarded `weight' is smaller than $10^{-5}$ in all our calculations. Consequently, we truncate the `initial' tensor to a size $16 \times 4 \times 16 \times 4$ after the initial imaginary time steps.

Furthermore, during the coarse graining process, we may do several imaginary time steps between each two space steps. Empirically we found that inside and close to the critical phase we have to do 2 time steps per 1 space step, while 1 time step per 1 space step suffices away from critical phase.}

\vspace{5mm}
\section{Conclusions}\label{conclu}

The XXZ model in a staggered magnetic field was previously studied by analytical methods for $\Delta=0$
and by exact diagonalization for $\Delta\ne 0$~\cite{Alcaraz_1995}. System sizes were 8-16 sites. Here,
we are able to study systems of up to 1024 sites  using a tensor network method. In this way it is possible to obtain the low lying spectrum with good precision. From the spectrum we derive the phase diagram as shown in Fig.~1.

The massless phase of the system is analyzed in considerable detail and compared to predictions of Luttinger liquid theory.
In particular, conformal towers predicted by this theory are compared to numerical data. Generally we find agreement, which suggests that the XXZ model in a staggered field resembles Luttinger liquid in the massless phase. However, close to the FM boundary we obtain qualitative differences which requires further investigation.
A definitive conclusion  needs more and, in particular, more precise numerical data.

Note that the results for the low-lying spectrum presented in section~\ref{sec-XXZ} and for the conformal towers presented in~\ref{sec-finite-size} are not produced in  separate calculations. At each parameter set $(\Delta, h_s)$
we only need one run of the HOTRG program to obtain all the data discussed in this paper.

\acknowledgments

M.V.R. acknowledges funding by the Deutsche Forschungsgemeinschaft (DFG, German Research Foundation) under Germany's Excellence Strategy – EXC-2123/1, and the support of Olle Engkvist Byggm\"astare Foundation under Decision 198-0389.

%


\end{document}